\documentclass[11pt]{JHEP3}
\usepackage{latexsym}

\renewcommand{\nc}{\newcommand}
\newcommand{\rnc}{\renewcommand}

\nc{\be}{\begin{equation}}
\nc{\ee}{\end{equation}}
\nc{\bea}{\begin{eqnarray}}
\nc{\eea}{\end{eqnarray}}

\nc{\trac}[2]{{\textstyle\frac{#1}{#2}}}

\nc{\ex}[1]{\mbox{e}^{\,\textstyle#1}}

\nc{\CC}{\mathbb{C}}
\nc{\HH}{\mathbb{H}}
\nc{\PP}{\mathbb{P}}
\nc{\RR}{\mathbb{R}}
\nc{\ZZ}{\mathbb{Z}}
\nc{\II}{\mathbb{I}}
\nc{\EE}{\mathbb{E}}

\rnc{\a}{\alpha}
\rnc{\b}{\beta}
\rnc{\d}{\delta}
\nc{\ga}{\gamma}
\nc{\la}{\lambda}
\nc{\f}{\phi}
\nc{\p}{\psi}
\nc{\e}{\eta}
\rnc{\c}{\chi}
\nc{\eps}{\epsilon}
\nc{\om}{\omega}
\nc{\Om}{\Omega}

\nc{\ad}{\mathop{\mbox{ad}}\nolimits}
\nc{\tr}{\mathop{\mbox{tr}}\nolimits}
\nc{\Tr}{\mathop{\mbox{Tr}}\nolimits}
\nc{\Det}{\mathop{\mbox{Det}}\nolimits}
\rnc{\det}{\mathop{\mbox{det}}\nolimits}
\nc{\rk}{\mathop{\mbox{rk}}\nolimits}
\nc{\del}{\partial}
\nc{\diag}{\mathop{\mbox{diag}}\nolimits}
\nc{\ra}{\rightarrow}
\nc{\Ra}{\Rightarrow}
\nc{\LRa}{\Leftrightarrow}
\nc{\lra}{\leftrightarrow}
\nc{\ot}{\otimes}
\rnc{\ss}{\subset}
\nc{\nul}{\noindent\underline}
\nc{\non}{\nonumber\\}
\nc{\mat}[4]{\left(\begin{array}{cc}#1&#2\\#3&#4\end{array}\right)}
\rnc{\lg}{\frak{g}}
\nc{\G}[3]{\Gamma^{#1}_{\;{#2}{#3}}}
\nc{\nam}{\nabla_{\mu}}
\nc{\nan}{\nabla_{\nu}}
\nc{\dx}{\dot{x}}
\nc{\dxl}{\dot{x}^{\la}}
\nc{\dxm}{\dot{x}^{\mu}}
\nc{\dxn}{\dot{x}^{\nu}}
\nc{\ddx}{\ddot{x}}
\nc{\ddxm}{\ddot{x}^{\mu}}
\nc{\ddxn}{\ddot{x}^{\nu}}
\nc{\dxi}{\dot{\xi}}
\nc{\ddxi}{\ddot{\xi}}

\title{Scalar Field Probes of Power-Law Space-Time Singularities}

\author{Matthias Blau, Denis Frank, Sebastian Weiss\\
Institut de Physique, Universit\'e de Neuch\^atel\\
Rue Breguet 1, CH-2000 Neuch\^atel, Switzerland}

\abstract{We analyse the effective potential of the scalar wave equation near
generic space-time singularities of power-law type (Szekeres-Iyer metrics)
and show that the effective potential exhibits a universal and scale
invariant leading inverse square behaviour $\sim x^{-2}$ in the ``tortoise
coordinate'' $x$ provided that the metrics satisfy the strict Dominant
Energy Condition (DEC).  This result parallels that obtained in \cite{bbop2}
for probes consisting of families of massless particles (null geodesic
deviation, a.k.a.\ the Penrose Limit). The detailed properties of the
scalar wave operator depend sensitively on the numerical coefficient
of the $x^{-2}$-term, and as one application we show that timelike
singularities satisfying the DEC are quantum mechanically singular in the
sense of the Horowitz-Marolf (essential self-adjointness) criterion. We
also comment on some related issues like the near-singularity behaviour
of the scalar fields permitted by the Friedrichs extension.}

\begin{document}

\section{Introduction}

The study of scalar field propagation in non-trivial curved (and possibly
singular) backgrounds is of fundamental importance in a variety of
contexts including quantum field theory in curved backgrounds, cosmology,
the stability and quasi-normal mode analysis of black hole metrics etc.

Typically, this is studied within the context of a particular metric
or class of metrics. For certain purposes, however, only the knowledge
of the leading behaviour of the metric near a horizon or the
singularity is required. In that case, one can attempt to work with a
general parametrisation of the metric near that locus and, in this way,
ascertain which features of the results that have been obtained previously
for particular metrics are special features of those metrics or valid
more generally.

In particular, practically all explicitly known metrics with singularities
are of ``power-law type'' \cite{SI} in a neighbourhood of the singularity
(instead of showing, say, some non-analytic behaviour). In the spherically
symmetric case, the leading behaviour of generic metrics with such
singularities of power-law type is captured by the 2-parameter family
\be
ds^2 = \eta x^p(- dx^2 + dy^2) + x^q d\Omega_d^2
\label{insim}
\ee
of Szekeres-Iyer metrics \cite{SI,CS,gpsi}. The singularity,
located in these coordinates at $x=0$, is timelike for $\eta=-1$ and
spacelike for $\eta=+1$. This class of metrics thus provides an ideal
laboratory for investigating the behaviour of particles, fields, strings,
\ldots in the vicinity of a generic singularity of this type.

A first investigation along these lines was performed in \cite{bbop1,bbop2}
in the context of the Penrose Limit, i.e.\ of probing a space-time via the
geodesic deviation of families of massless particles. There it was shown that
the plane wave Penrose limits,
\be
ds^2 = g_{\mu\nu}dx^{\mu}dx^{\nu} \ra
2dudv + A_{ab}(u)x^a x^b du^2 + d\vec{x}^2\;\;,
\label{impl}
\ee
of metrics with singularities of power-law type have a universal
$u^{-2}$-behaviour near the singularity, $A_{ab}(u) \sim u^{-2}$, provided
that the near-singularity stress-energy (Einstein) tensor satisfies the
strict dominant energy condition (DEC). This behaviour, which is precisely
such that it renders the plane wave metric scale invariant \cite{mmhom},
had previously been observed in various particular examples and is thus
now understood to be a general feature of this class of singularities.

It is then natural to wonder whether a similar universality result can be
established in other circumstances or for other kinds of probes and if,
analogously, some energy condition plays a role in establishing this.
If one considers e.g.\ the Klein-Gordon equation $\Box\phi = 0$ for scalar
fields, it is not difficult to see \cite{mbmo,sdss} that
under certain conditions the scalar effective potential
$V_{\mathrm{eff}}$ for general metrics with singularities of power-law
type displays an inverse square behaviour, $V_{\mathrm{eff}}(x)
\sim x^{-2}$, near the singularity. 
This observation was then used in \cite{sdss} to study
the quasi-normal modes for black holes with generic singularities of
this type.

The purpose of this note is to study other aspects and consequences of
this universality. In particular, we will first show that the results
obtained in \cite{bbop2}, namely the scale invariant inverse square
behaviour of the wave profile $A_{ab}(u)$, as well as a crucial
\cite{prt,mmhom} lower bound on
the coefficients, have a precise and rather striking analogue in the
case of a scalar field. Schematically, this analogy can be expressed 
as
\be
\mathrm{strict\;DEC}\;\;\;\;\Ra\;\;\;\;
\left\{\begin{array}{lll} 
A_{ab}(u) \ra c_a \d_{ab} u^{-2} &
\mathrm{scale\;invariant} & (c_a \geq -1/4) \\
V_{\mathrm{eff}}(x) \ra c x^{-2} &
\mathrm{scale\;invariant} & (c \geq -1/4)
\end{array}
\right.
\label{summary}
\ee
Once again this shows that this inverse square behaviour, that had been
observed before in various specific examples in a variety of contexts,
is a general feature of a large class of space-time singularities.
The precise statements are derived in sections 2.2 and 2.3 and discussed
in section 2.4, while sections 2.5 and 2.6 deal with minor variations
of this theme.

We hasten to add that if such an inverse square behaviour were universally
true without any further qualifications then it would probably have to
be true on rather trivial (dimensional) grounds alone. What makes the
results obtained here and in \cite{bbop2} more interesting is that a
priori in either case a more singular behaviour can and does occur and
is only excluded provided that some further (e.g.\ positive energy)
condition is imposed.

The significance of the $x^{-2}$-behaviour is that (as anticipated in
(\ref{summary})), the corresponding Schr\"odinger operator $-\del_x^2 +
cx^{-2}$, to which we will have reduced the Klein-Gordon operator, defines
a scale invariant ($c$ is dimensionless) ``conformal quantum mechanics''
\cite{aff} problem. Thus, here and in \cite{bbop2} we find a rather
surprising emergence of scale invariance in the near-singularity limit.
One implication of this scale invariance in the plane wave case, discussed
in \cite{mmm}, is that it leads to a Hagedorn-like behaviour of string
theory in this class of backgrounds that is quite distinct from that in
plane wave backgrounds with, say, a constant profile and more akin to
that in Minkowski space. It would be interesting to explore other
consequences of this near-singularity scale invariance.

This class of scale invariant models has recently also appeared and been
discussed in various other related settings, most notably in the analysis
of the near-horizon (rather than the near-singularity) properties of
black holes, see e.g.\ \cite{cdkkpt,ggpkt,gsvcqm,bip,crcqm1},  where
the emergence of scale invariance can largely be attributed to the
near-horizon AdS geometry, as well as in quantum cosmology \cite{boris}.

Having reduced the Klein-Gordon operator to the Schr\"odinger operator
$-\del_x^2 + cx^{-2}$ (after a separation of variables and a unitary
transformation), one can then turn to a more detailed spectroscopy
of the Szekeres-Iyer metrics by analysing the properties of this
operator. Indeed, as is well known, the inverse square potential is a
critical borderline case in the sense that the functional analytic properties
of this operator depend in a delicate way on the numerical value of the
coefficient $c$. This value, in turn, depends on the dimension $d$ (number
of transverse dimensions) and the Szekeres-Iyer parameter $q$ (it turns
out to be independent of $p$, while the corresponding coefficients $c_a$
in the Penrose limit case typically depend on $(p,q)$ and $d$).

As one application, we will analyse the Horowitz-Marolf criterion
\cite{hm} for general singularities of power-law type.  Horowitz and
Marolf defined a static space-time to be quantum mechanically non-singular
(with respect to a certain class of test fields) if the evolution of
a probe wave packet is uniquely determined by the initial wave packet
(as would be the case in a globally hyperbolic space-time) without
having to specify boundary conditions at the classical singularity. This
criterion can be rephrased as the condition that the (spatial part of
the) Klein-Gordon operator be essentially self-adjoint (and thus have
a unique self-adjoint extension).

While such a necessarily only semi-classical analysis is certainly not
a substitute for a full quantum gravitational analysis, it nevertheless
has its virtues since one can learn what kind of problems persist, can
arise or can be resolved when passing from test particles to test fields.

Intuitively one might expect a classical singularity with a sufficiently
``positive'' (in an appropriate sense) matter content to remain singular
even when probed by non-stringy test objects other than classical
point particles.  This line of thought was one of the motivations for
analysing the Horowitz-Marolf criterion in this framework, and we will
indeed be able to show (section 3.4) that

\begin{quote}
metrics with timelike singularities of power-law type satisfying the strict
Dominant Energy Condition remain singular when probed with scalar waves.
\end{quote}

A second issue we will briefly address is that of the allowed
near-singularity behaviour of the scalar fields for a given self-adjoint
extension (section 3.5). A priori, one might perhaps expect a sufficiently
repulsive singularity to be regular in the Horowitz-Marolf sense simply
because the corresponding unique self-adjoint extension forces the scalar
field to be zero at the singularity, thus in a sense again excluding
the singularity from the space-time. It is also possible, however, and
potentially more interesting, to have a self-adjoint extension with
scalar fields that actually probe the singularity in the sense that
they are allowed to take on non-zero values there. We propose to call
such singularities ``hospitable'', establish once again a relation,
albeit not a strict correlation, with the DEC, and show among other
things that, in a suitable sense, half of the Horowitz-Marolf regular
power-law singularities are hospitable whereas the others are not.

\section{Universality of the Effective Scalar Potential for 
Power-Law Singularities}

\subsection{Geometric Set-Up}

Even though we will ultimately be interested in the properties of
the scalar wave (Klein-Gordon) equation $(\Box-m^2)\phi = 0$ in
the Szekeres-Iyer metrics (\ref{insim}), to set the stage it will be
convenient to begin the discussion in the more general setting of metrics
with a hypersurface orthogonal Killing vector.  The general set-up
here and in section 3.1 is modelled on the approach of \cite{wald1}
(with minor adaptations to allow for both timelike and
spacelike singularities).

We begin with the $n$-dimensional metric
\be
ds^2 =  \eta a^2 dy^2 + h_{ij}dx^i dx^j
\label{met1}
\ee
where $a$ and $h_{ij}$ are independent of $y$, $\xi = \del_y$ is a
hypersurface orthogonal Killing vector with norm $\xi^{\mu}\xi_{\mu}
= \eta a^2$, and thus timelike (spacelike) for $\eta=-1$ ($\eta=+1$).
Correspondingly we assume that the metric $h_{ij}$ induced on the
hypersurfaces $\Sigma_y \cong \Sigma$ of constant $y$ 
is Riemannian (Lorentzian) for $\eta=-1$
($\eta=+1$).

Denoting the covariant derivatives with respect to the metric
$h_{ij}$ by $D_i$, the wave operator
\be
\Box\equiv\frac{1}{\sqrt{-\det g}}\del_{\mu}\sqrt{-\det g}g^{\mu\nu}\del_{\nu}
\ee 
is easily seen to take the form
\be
\Box = a^{-2} (\eta\del_y^2 + a D^i a D_i)\;\;.
\ee
Thus the massive wave equation $(\Box - m^2)\phi=0$ can be written as
\be
\del_y^2 \phi = - A \phi\;\;,
\ee
where $A$ is the operator
\be
A =  \eta a D^i a D_i - \eta a^2 m^2\;\;.
\label{A1}
\ee

Assuming now spherical symmetry, the metric takes the warped product form
\be
ds^2 = \eta a(x)^2 dy^2 - \eta b(x)^2 dx^2 + c(x)^2 d\Omega_{d}^2
\label{ssm1}
\ee
where $d\Omega_d^2$, $d=n-2$, denotes the standard metric on the 
$d$-sphere $S^d$. It will be apparent from the following that  
the assumption of spherical symmetry could be relaxed - we will only
use the warped product form of the metric in an essential way.

We could fix the residual $x$-reparametrisation 
invariance by introducing the ``area radius'' $r=c(x)$ as a new coordinate.
However, for the following it will be more convenient to choose the
gauge $a(x)=b(x)$ (i.e.\ $x$ is a ``tortoise coordinate''
for $\eta=-1$ respectively ``conformal time'' for $\eta = +1$),
\be
ds^2 = \eta a(x)^2 (-dx^2 + dy^2) + c(x)^2 d\Omega_{d}^2\;\;.
\label{ssm2}
\ee
Then the operator $A$ is
\be
A = - \sigma^{-1} \del_x \sigma\del_x +\eta a^2 c^{-2} \Delta_d - \eta a^2
m^2\;\;,
\label{A2}
\ee
where $\sigma(x) = c(x)^d$ and $\Delta_d$ denotes the Laplacian on $S^d$.

To put $A$ into standard Schr\"odinger form, we transform from
the functions $\phi(x)$ to the half-densities (cf.\ (\ref{hd1}))
$\tilde{\phi}(x) = \sigma^{1/2}\phi(x)$. The corresponding
unitarily transformed operator $\tilde{A}$ is
\bea
\tilde{A} &=& \sigma^{1/2}A\sigma^{-1/2} = -\del_x^2  + V 
+\eta a^2 c^{-2}\Delta_d - \eta a^2 m^2\non
V(x) &=& \sigma(x)^{-1/2} (\del_x^2 \sigma(x)^{1/2})\;\;.
\label{V}
\eea
After the usual separation of variables in the $y$-direction,
\be
\tilde{\phi}(y,x,\theta^a) = \ex{-iE y}\;\tilde{\phi}(x,\theta^a)\;\;,
\ee
and the decomposition into angular spherical harmonics
$Y_{\ell\vec{m}}(\theta^a)$, with
\bea
-\Delta_d Y_{\ell\vec{m}}(\theta^a)&=& \ell_d^2 Y_{\ell\vec{m}}(\theta^a)\non
\ell_d^2 &=& \ell (\ell + d -1) \;\;,
\eea
the Klein-Gordon equation for the metric (\ref{ssm2})
reduces to a standard one-dimensional time-independent 
Schr\"odinger equation 
\be
\left[-\del_x^2 +
V_{\mathrm{eff},\ell}(x)\right]\tilde{\phi}(x) = E^2 \tilde{\phi}(x)
\ee
($\tilde{\phi}(x) = \tilde{\phi}_{E,\ell,\vec{m}}(x)$)
with effective scalar potential 
\be
V_{\mathrm{eff},\ell}(x) = V(x) - \eta a(x)^2(\ell_d^2 c(x)^{-2}  + m^2)
\label{Veff1}
\;\;.
\ee

\subsection{The Effective Scalar Potential for Power-Law Singularities}

The leading behaviour of generic (spherically symmetric)
metrics with singularities of power-law type\footnote{Such metrics encompass
practically all
explicitly known singular spherically symmetric solutions of the Einstein
equations like the Lema\^itre-Tolman-Bondi
dust solutions, cosmological singularities of the Lifshitz-Khalatnikov
type, etc. On the
other hand, this class of metrics does prominently {\em not} include
the BKL metrics \cite{BKL} describing the chaotic oscillatory approach
to a spacelike singularity. Whether or not such a behaviour occurs depends
in a delicate way on the matter content, see e.g.\ \cite{dhn} and references
therein.}, i.e.\ metrics of the general form \cite{SI}
\be
ds^2 = -dt^2 + [t-\tau(r)]^{2a} f(r,t)^2 dr^2 +
[t-\tau(r)]^{2b}g(r,t)^2d\Omega_d^2\;\;,
\ee
with $f$ and $g$ functions of $r$ and $t$ that are regular and non-vanishing
at the location $t=\tau(r)$ of the singularity,
is captured by the
2-parameter family of Szekeres-Iyer metrics \cite{SI,CS} (see also
\cite{bbop2} and the generalisation to string theory backgrounds
discussed in \cite{gpsi})
\be
ds^2 = \eta x^p(- dx^2 + dy^2) + x^q d\Omega_d^2 \;\;.
\label{sim2}
\ee
The Kasner-like exponents $p,q \in \RR$ characterise the behaviour of the 
geometry near the singularity at $x=0$. 
This singularity is timelike for $\eta=-1$ 
($x$ is a radial coordinate) and spacelike for $\eta=+1$ (with $x$ 
a time coordinate). In particular, these metrics possess the hypersurface
orthogonal Killing vector $\del_y$, and 
are already in the ``tortoise'' form (\ref{ssm2}), with $a(x)^2 = x^p$ and
$c(x)^2=x^q$. Thus we can directly read off the effective scalar potential
from the results of the previous section.

From (\ref{V}), we deduce, with $\sigma(x) = x^{dq/2}$, that
\be
V(x)= s(s-1) x^{-2} \;\;\;\;\;\; s = \frac{dq}{4}\;\;.
\ee
Thus, from (\ref{Veff1}) we find (see also \cite{sdss})
\be
V_{\mathrm{eff},\ell}(x)=s(s-1)x^{-2} - \eta \ell_d^2 x^{p-q} - \eta m^2 x^p
\label{siveff}
\ee
We are interested in the leading behaviour of this potential as $x \ra 0$
(subdominant terms can in any case not be trusted as we have only kept the 
leading terms in the metric (\ref{sim2})). For the time being
we will consider the massless case $m^2=0$ (see section 2.5 for $m^2 \neq
0$).

Provided that $s(s-1)\neq 0$, which term in (\ref{siveff}) dominates
depends on $p$ and $q$. When $q < p+2$, one finds
\be
q < p+2: \;\;\;\;\;\;V_{\mathrm{eff},\ell}(x) \ra s(s-1) x^{-2} \;\;.
\label{case1}
\ee
The two salient features of this potential are the inverse square behaviour
and a coefficient $c$ that is bounded from below by $-1/4$, 
\be
c=s(s-1) \geq -\frac{1}{4}\;\;,
\ee
with equality for $s=1/2$, i.e.\ $q=2/d$. 

As mentioned in the introduction, the significance of the
$x^{-2}$-behaviour is that it defines a scale invariant ``conformal
quantum mechanics'' \cite{aff} problem, discussed more recently in
related contexts e.g.\ in \cite{cdkkpt,ggpkt,gsvcqm,bip,crcqm1,boris}.
Moreover, for practical purposes \cite{sdss,gkos} the virtue of the
$x^{-2}$ (as opposed to a more singular) behaviour is that it leads to
a standard regular-singular differential operator.

The significance of the bound on $c$ is that in this range the operator 
$-\del_x^2 + c/x^2$ is positive, as can be seen by writing 
\be
-\del_x^2 + s(s-1) x^{-2} = (\del_x + sx^{-1})(-\del_x + s x^{-1}) 
=(-\del_x + sx^{-1})^{\dagger}(-\del_x + s x^{-1})\;\;.
\ee 

When $q=p+2$, the metric is conformally flat, 
both terms in (\ref{siveff}) contribute equally, and one
again finds the $x^{-2}$-behaviour
\be
q = p+2: \;\;\;\;\;\;V_{\mathrm{eff},\ell}(x) \ra c x^{-2}\;\;,
\ee
where now
\be 
c= s(s-1) - \eta \ell_d^2\;\;.
\ee
Thus in this case $c$ is still bounded by $-1/4$ for timelike singularities,
while $c$ can become arbitrarily negative for sufficiently large values of
$\ell_d^2$ for $\eta = +1$.

Once $q > p+2$, the second term in (\ref{siveff}) dominates (for $\ell_d^2
\neq 0$), and one finds the more singular leading behaviour
\be
q > p+2: \;\;\;\;\;\;
V_{\mathrm{eff},\ell}(x) \ra -\eta \ell_d^2\; x^{-2-a}\;\;\;\;
a > 0 \;\;.
\label{sveff}
\ee

Examples of metrics with $q\leq p+2$ are the Schwarzschild and
Friedmann-Robertson-Walker (FRW) metrics and indeed, as we will recall below,
all metrics satisfying the strict Dominant Energy Condition.

In particular, for the $(d+2)$-dimensional (positive or negative mass)
Schwarzschild metric, one has
\be
\mathrm{Schwarzschild:}\;\;\;\;\;\;p=\frac{1-d}{d} \;\;\;\;\;\; q =
\frac{2}{d}\;\;,
\label{sspq}
\ee
as is readily seen by expanding the metric near the singularity and
going to tortoise coordinates. Thus the Schwarzschild metric has $s=1/2$
and $c$ takes on the $d$-independent extremal value $c=-1/4$, as observed 
before e.g.\ in \cite{gkos,sdss} in related contexts.

For decelerating cosmological FRW metrics, with cosmological
scale factor (in comoving time) $\sim t^h$, $0 < h < 1$,
\be
h=\frac{2}{(d+1)(1+w)}\;\;,
\label{frwh}
\ee
with $w$ the equation of state parameter, $P=w\rho$, one finds 
\cite{bbop1,bbop2}
\be
\mathrm{FRW:}\;\;\;\;\;\; p=q=\frac{2h}{1-h}\;\;,
\label{frwpq}
\ee
as can be seen by going to conformal time. A routine calculation shows
that the above result (\ref{case1}) for the purely $x$-dependent part of
the effective potential (with $x$
conformal time) is actually an exact result, and not an artefact of the
near-singularity Szekeres-Iyer approximation.

It remains to discuss the case when $q < p+2$, so that the first term
in (\ref{siveff}) would be dominant, but the coefficient $s(s-1)=0$.
When $s=0$, then one has $q=0$ and this is generally interpreted \cite{SI}
as corresponding not to a true central singularity (as the radius of the
transverse sphere remains constant as $x\ra 0$) but as a shell crossing
singularity. 

The other possibility is $s=1$, i.e.\ $q=4/d$. This is a case in which
(because of the cancellation of the leading terms) subleading corrections to
the metric (\ref{sim2}) can become relevant and should be retained.  
An example of metrics with $s=1$ is provided by FRW metrics with a
radiative equation of state. Using (\ref{frwpq}), one has
\be
q = \frac{4}{d} \LRa h = \frac{2}{d+2} \LRa w = \frac{1}{d+1}\;\;,
\ee
which is precisely the equation of state parameter for radiation. However,
as follows from the remark above, in this special case the vanishing of 
the effective potential for $p=q$ is actually an exact result.

\subsection{The Significance of the (Strict) Dominant Energy Condition}

We have seen that generically the leading behaviour of the scalar
effective potential near a singularity of power-law type is 
either $\sim x^{-2}$ or $\sim x^{p-q}$. We will now recall from
\cite{SI,bbop2} that 
the latter behaviour can arise only for metrics violating the
strict Dominant Energy Condition (DEC). While there is nothing
particularly sacrosanct about the DEC, and other energy conditions
could be considered, the DEC appears to play a privileged role in 
exploring and understanding the $(p,q)$-plane of Szekeres-Iyer metrics.

The \textit{Dominant Energy Condition} on the stress-energy tensor
$T^{\mu}_{\;\nu}$ (or Einstein tensor $G^{\mu}_{\;\nu}$) \cite{HE}
requires that for every timelike vector $v^\mu$, $T_{\mu\nu}v^{\mu}v^{\nu}
\geq 0$, and $T^{\mu}_{\;\nu}v^{\nu}$ be a non-spacelike vector. This
may be interpreted as saying that for any observer the local energy
density is non-negative and the energy flux causal.

The Einstein tensor of Szekeres-Iyer metrics is diagonal, hence so is
the corresponding stress-energy tensor. In this case, the DEC reduces to
\be
\rho \geq |P_{i}| \;\;,
\label{dec}
\ee
where $-\rho$ and $P_i$, $i=1,\ldots,d+1$ are the timelike and spacelike
eigenvalues of $T^{\mu}_{\;\nu}$ respectively.
We say that the {\em strict\/} DEC is satisfied if these are strict
inequalities and we will say that the matter content (or equation of
state) is ``extremal'' if at least one of the inequalities is saturated.

Now it follows from the explicit expression for the components
\bea
G^x_x 
&=& -\trac{1}{2}d(d-1) x^{-q} - \trac{1}{8}\eta d q
((d-1)q+2p)x^{-(p+2)}    \non
G^y_y 
&=&     -\trac{1}{2}d(d-1) x^{-q} + \trac{1}{8}\eta d q (2p + 4 -
(d+1)q) x^{-(p+2)}    
\label{gsi}
\eea
of the Einstein tensor that 
for $q > p+2$ the relation between $-\rho$ and the radial pressure 
$P_r$ (identified with $G_x^x$ and $G_y^y$ - which is which depends on the
sign of $\eta$) becomes extremal as $x\ra 0$ \cite{SI,bbop2},
\be
q > p+2:\;\;\;\;\;\;
G_x^x-G^y_y \ra 0 \;\;\;\;\LRa\;\;\;\;\rho + P_r \ra 0\;\;.
\ee
Put differently, $q \leq p+2$ is a necessary condition for the strict DEC
to hold, and thus for metrics satisfying the strict DEC the leading behaviour
of the effective potential is always $V_{\mathrm{eff},\ell}(x)\ra c x^{-2}$.

As an aside, we note that it follows from (\ref{gsi}) that precisely
those metrics that satisfy the physically more reasonable (non-negative
pressure) and more common extremal near-singularity equation of state
$\rho = + P_r$ have $q=2/d$, i.e.\ $s=1/2$, leading to the critical value
$c=-1/4$ frequently found in applications (to e.g.\ Schwarzschild-like
geometries).

\subsection{Comparison with Massless Point Particle Probes (the Penrose Limit)}

In the previous section we have established that 
\begin{enumerate}
\item
for metrics with singularities of power-law type satisfying the strict
DEC the leading behaviour of the scalar effective potential near the
singularity is 
\be
V_{\mathrm{eff},\ell}(x) \ra c x^{-2}
\ee
\item this class of potentials is singled out by its scale invariance;
\item
the corresponding coefficient $c$ of the effective potential is bounded
from below by $-1/4$ unless one is on the border to an extremal equation
of state. 
\end{enumerate}

These observations bear a striking resemblance to the
results obtained recently in \cite{bbop2} in the study of plane wave
Penrose limits
\be
ds^2 = g_{\mu\nu}dx^{\mu}dx^{\nu} \ra
2dudv + A_{ab}(u)x^a x^b du^2 + d\vec{x}^2\;\;,
\label{pwm}
\ee
of space-time singularities. Namely, it was shown in \cite{bbop2} that 
\begin{enumerate}
\item
Penrose limits of metrics with singularities of power-law type
show a universal $u^{-2}$-behaviour near the singularity,
\be
A_{ab}(u) \ra c_a \d_{ab} u^{-2}\;\;,
\label{2}
\ee
provided that the strict DEC is satisfied; 
\item such plane waves are singled out \cite{mmhom} by their scale
invariance, reflected e.g.\ in the isometry $(u,v) \ra (\lambda u,
\lambda^{-1}v)$ of the metric (\ref{pwm}, \ref{2});
\item
the coefficients $c_a$ (related to the harmonic oscillator 
frequency squares by $c_a=-\omega_a^2$) are bounded from below by
$-1/4$ unless one is on the border to an extremal equation of 
state.\footnote{One significance of this bound on the $c_a$ is that 
in this range one can consider the possibility to extend the string modes
across the singularity at $u=0$ \cite{prt}.}
\end{enumerate}

The similarity of these two sets of statements is quite remarkable because
the objects these statements are made about are rather different. For
example, the potential is that of a one-dimensional motion on the half
line in one case, and that of a $d$-dimensional harmonic oscillator
(with time-dependent frequencies) in the other.

The analogy with the above statements about scalar effective potentials
is brought out even more clearly if one reinterprets \cite{bbop1,bbop2}
the Penrose limit in terms of null geodesic deviation in the original
space-time. Then this result can be rephrased as the statement that
the leading behaviour of the geometry as probed by a family of massless
point particles near a singularity is that of a plane wave with a $u^{-2}$
geodesic effective potential. The analogy with the results of the
previous section should now be apparent.

One minor difference between the results obtained here and those of
\cite{bbop2} is that in the case of Penrose limits the strict DEC
needed to be invoked only in the case of spacelike singularities,
$\eta=+1$, timelike singularities always giving rise to plane waves
with a $u^{-2}$-behaviour. This should be regarded as an indication
(cf.\ the discussion in \cite[Section 4.4]{bbop2}) that scalar waves
are better probes of timelike singularities than massless point particles.

\subsection{Massive Scalar Fields and Geodesic Incompleteness}

The simple above analysis can evidently be generalised in various
ways, e.g.\ by considering other kinds of probes. We will briefly
comment on the two most immediate generalisations, namely massive and
non-minimally coupled scalar fields.

We begin with a massive scalar for which the effective potential is
\be
V_{\mathrm{eff},\ell}(x)=s(s-1)x^{-2} - \eta \ell_d^2 x^{p-q} - \eta m^2 x^p
\label{siveff2}
\ee
For the mass term to be relevant (dominant) as $x\ra 0$ it is clearly
necessary that $p < -2$ and $q<0$. Intuitively one might expect a
mass term to be irrelevant at short distances near a singularity. This
expectation is indeed borne out: as we will now show, for metrics
satisfying the above inequalities the would-be singularity at $x=0$ is
at infinite affine distance for causal geodesics so that such space-times
are actually causally geodesically complete.

Null geodesics were analysed in \cite{bbop2}. Here we
generalise this to causal geodesics. 
In terms of the conserved angular and $y$-momentum $L$ and $P$, the geodesic
equation for the metric (\ref{sim2}) reduces to
\be
\dot{x}^2 = P^2 x^{-2p} + \eta L^2 x^{-p-q} +\eta\epsilon x^{-p}\;\;,
\label{sig}
\ee
where $\epsilon = 0$ ($\epsilon = 1$) for null (timelike) geodesics
respectively. 

For $\eta=-1$, if the first term in (\ref{sig}) is sub-dominant the geodesic
effective potential is repulsive (e.g.\ via the angular momentum barrier) and
the geodesics will not reach $x=0$. Thus generic timelike geodesics will
reach $x=0$ only if $(p,q)$ lie in the positive wedge bounded by the lines
$p=0$ and $p=q$. Radial null
geodesics do not feel any repulsive force, and solving 
\be
\dot{x}^2 \sim x^{-2p}\;\;\Ra \;\; x(u) \sim \left\{\begin{array}{cc}
 u^{1/(p+1)} & p\neq -1\\
 \exp u     & p = -1
\end{array}\right.
\ee
shows that $x=0$ is reached at a finite value of the affine
parameter only for $p >-1$. We thus conclude that Szekeres-Iyer 
metrics with $\eta = -1$ and 
$p \leq -1$ are causally geodesically complete. 
In particular, therefore, the mass term in the scalar effective potential
is sub-dominant for metrics with honest timelike power-law singularities, and 
for all such metrics
the scalar effective potential has the same leading behaviour as in the
massless case.

For $\eta =+1$, the situation is more complex as all three terms in
(\ref{sig}) are positive. If the first term dominates, either because
of suitable inequalities satisfied by $(p,q)$ or, for any $(p,q)$,
because one is considering radial null geodesics, the analysis and
conclusions are identical to the above. Namely, $x=0$ is at finite affine
distance for $p > -1$. Analogously, if the second term dominates (e.g.\
for angular null geodesics) one finds the condition $p+q > -2$, and
if the third term dominates one has $p > -2$. Since one needs $p < -2$
for the mass term to dominate in the scalar effective potential, 
only the second case is possible. But then
the condition $p + q > -2$, with $p <-2$, implies $q > 0$, so that the
angular momentum term in the effective potential dominates the mass term.

We thus conclude that, for both $\eta=+1$ and $\eta=-1$, the mass term is
always subdominant for metrics that are causally geodesically incomplete
at $x=0$.

As an aside we note that the Szekeres-Iyer metrics for which the mass
term does dominate ($p<-2$ and $q<0$), in addition to being non-singular,
also necessarily violate the strict DEC.

\subsection{Non-Minimally Coupled Scalar Fields}

We will now very briefly also consider a non-minimally coupled scalar field
\be
(\Box - \xi R)\phi = 0\;\;.
\ee
The Ricci scalar of the Szekeres-Iyer metric (\ref{sim2}) is
\be
R = d(d-1) x^{-q} - \trac{1}{4}\eta (4p + 4qd - d(d+1) q^2) x^{-(p+2)}\;\;,
\ee
where once again only the leading order term should be trusted and retained.
Thus the new effective potential
\be
V_{\mathrm{eff},\ell}^\xi(x) = V_{\mathrm{eff},\ell}(x) - \eta \xi x^p R
\ee
is again a sum of two terms, proportional to $x^{-2}$ and $x^{p-q}$
respectively, so that the dominant behaviour is still $\sim x^{-2}$ 
provided that the metric does not violate the strict DEC. For $q < p+2$
and the conformally invariant coupling
\be
\xi = \xi_* = \frac{d}{4(d+1)}\;\;,
\ee
one finds 
\be
V_{\mathrm{eff},\ell}^{\xi_*}(x) 
= \frac{(p-q)d}{4(d+1)}x^{-2} =(p-q)\xi_* x^{-2} \;\;.
\ee
Note that with this conformally invariant coupling the coefficient
$c$ now depends on $p-q$ rather than on $q$. The appearance of $(p-q)$
could have been anticipated since for $p=q$ the Szekeres-Iyer metric is
conformal to an $x$-independent metric, and hence a conformal coupling
cannot generate an $x$-dependent effective potential. Note also that
for the conformal coupling (and, indeed, generic values of $\xi$) the
coefficient $c$ is no longer bounded by $-1/4$ so that the Schr\"odinger
operator is no longer necessarily bounded from below.

\section{Self-Adjoint Physics of Power-Law Singularities}

In the previous section we have determined the leading behaviour of the
scalar wave operators near a power-law singularity. In this section we
will now study various aspects of these operators. 

\subsection{Functional Analysis Set-Up}

In order to analyse the properties of the wave operator, we will need to
equip the space of scalar fields with a Hilbert space structure. We will
be pragmatic about this and introduce the minimum amount of structure
necessary to be able to say anything of substance. 

We thus return to the discussion of section 2.1, now being more specific 
about the spaces of functions the various operators appearing there act 
on \cite{wald1}, beginning with the operator $A$ introduced in (\ref{A1}), 
\be
A =  \eta a D^i a D_i - \eta a^2 m^2\;\;.
\ee
Since $D^i D_i$ is symmetric (formally self-adjoint) with respect to
the natural spatial density $\sqrt{-\eta \det h}$ induced on the slices
$\Sigma$ of constant $y$ by the metric (\ref{met1}), the operator $A$ is
symmetric with respect to the scalar product
\bea
(\phi_1,\phi_2) &=& \int d^{n-1}x\; \sigma \phi_1^*\phi_2\non
\sigma &=& a^{-1} \sqrt{-\eta \det h} =  \eta\sqrt{-\det g} g^{yy}\;\;,
\label{sp1}
\eea
on $D(A) = C_0^{\infty}(\Sigma)$, 
\be
(A\phi_1,\phi_2) = (\phi_1,A\phi_2)\;\;.
\ee
Moreover, for $\eta=-1$ the operator $A$ is positive, 
\be
\eta = -1 \;\;\Ra\;\; (\phi,A\phi)\geq 0 \;\;.
\ee
We are thus led to introduce the Hilbert space $L^2(\Sigma,\sigma
d^{n-1}x)$ of functions on $\Sigma$ square integrable with respect to
the above scalar product.
 
Passing to spherically symmetric metrics (\ref{ssm1}) in the tortoise
gauge (\ref{ssm2}), $A$ takes the form (\ref{A2})
\be
A = - \sigma^{-1} \del_x \sigma\del_x +\eta a^2 c^{-2} \Delta_d - \eta a^2
m^2\;\;,
\ee
where $\sigma(x) = c(x)^d$. Since $A$ is symmetric with respect to
the scalar product (\ref{sp1}), the unitarily transformed operator
\be
\tilde{A} = \sigma^{1/2}A\sigma^{-1/2}\;\;, 
\ee
acting on the half-densities 
\be
\tilde{\phi}(x) = \sigma(x)^{1/2}\phi(x)\;\;,
\ee
is symmetric
with respect to the corresponding ``flat'' ($\sigma(x) \ra 1$) scalar product
\be
<\tilde{\phi}_1,\tilde{\phi}_2> := \int dx\,d\Omega\; \tilde{\phi}_1^*
\tilde{\phi}_2 = (\phi_1,\phi_2)\;\;.
\label{hd1}
\ee
We now assume that the metric develops a singularity at some value of $x$,
where e.g.\ the area radius goes to zero, $r\equiv c(x) \ra 0$,
which we may as well choose to happen at $x=0$. Thus we consider 
$x\in (0,\infty)$ and take $\Sigma = \RR^{n-1}\backslash \{0\}$, 
parametrised by $x$ and the angular coordinates. 

Then the initial domain of $\tilde{A}$ is
$D(\tilde{A}) = C_0^{\infty}(\RR^{n-1}\backslash\{0\})$ 
or $\tilde{D}(\tilde{A}) = C_0^{\infty}(\RR_+)\otimes
C^{\infty}(S^d)$, which are dense in the unitarily transformed 
Hilbert space 
\be
L^2(\RR^{n-1}\backslash\{0\}, dx\,d\Omega) \cong 
L^2(\RR_+,dx)\otimes L^2(S^d,d\Omega)\;\;.
\ee 
Decomposing the second factor
into eigenspaces of the Laplacian $\Delta_d$ on $S^d$,
\be
L^2(\RR_+,dx)\otimes L^2(S^d,d\Omega)=\bigoplus^\infty_{\ell=0}L_\ell\;\;,
\ee
and defining $\tilde{D}_{\ell} = \tilde{D} \cap L_{\ell}$, one has
\be
\tilde{A}|_{\tilde{D}_{\ell}}= \tilde{A}_{\ell} \otimes \II\;\;,
\ee
where
\be
\tilde{A}_{\ell} = -\del_x^2 + V_{\mathrm{eff},\ell}(x) 
\ee
with $V_{\mathrm{eff},\ell}(x)$ given in (\ref{Veff1}).

Questions about the original operator $A$ can thus be reduced to questions
about the family $\{\tilde{A}_{\ell}\}$ of standard Schr\"odinger-type
operators. For example, to show that $A$ is essentially self-adjoint on
$D(A)$ it is sufficient to prove that, for each $\ell$, $\tilde{A}_{\ell}$
is essentially self-adjoint on $C_0^{\infty}(\RR_+)$.

While one can analyse this question of self-adjointness just as readily
for $\eta =+1$ as for $\eta=-1$, the physical significance of this
condition in the case of spacelike singularities is not clear to us. Thus
we will focus on static space-times with timelike singularities in the
following and set $\eta=-1$. An extension of the general formalism to
stationary non-static space-times is developed in \cite{seggev}.

We conclude this section with a comment on the choice of Hilbert space 
structure. The $L^2$ Hilbert space introduced above is certainly a 
natural choice, but not the only one possible. Based on physical requirements
such as the finiteness of the energy of scalar field probes, 
other (Sobolev) Hilbert space structures
have been proposed in the literature - see e.g.\ \cite{ishihoso,stalker}.
The energy is, by definition, 
\be
E[\phi] = \int_{\Sigma} \sqrt{h}d^{n-1}x\;T_{\mu\nu}(\phi)\xi^\mu n^\nu\;\;,
\ee
where $T_{\mu\nu}(\phi)$ is the stress energy tensor of the scalar field,
$\xi = \del_y$ is the timelike Killing vector, and $n$ the unit normal
to $\Sigma$. In the present case this reduces to
\be
E[\phi] = \int_{\Sigma} \sigma d^{n-1}x\; T_{yy}\;\;,
\label{ef1}
\ee
which identifies $T_{yy}$ as the energy density with respect to the measure
$\sigma d^{n-1}x$ employed above \cite{stalker}. For a minimally coupled
complex scalar field one has 
\be
T_{yy} = \frac{1}{2}\left[\del_y\phi^*\del_y \phi + a^2
h^{ij}\del_i\phi^*\del_j\phi\right]\;\;.
\ee
Thus, with an integration by parts (certainly allowed for $\phi \in D(A)$) 
the energy can be written as
\bea
E[\phi] &=& \int_{\Sigma} \sigma d^{n-1}x\;\left(\del_y \phi^*\del_y\phi +
\phi^*A\phi\right)\non
&=& (\del_y\phi,\del_y\phi) + (\phi,A\phi)\;\;.
\label{ef2}
\eea 
For a comparison of the two definitions (\ref{ef1}) and (\ref{ef2}) of the
energy and the role of boundary terms, see e.g.\ the discussion in
\cite{gibharish} and the comment in section 3.5 below.
Adopting the expression (\ref{ef2}) as the definition of the energy 
suggests introducing a Sobolev structure on the space of scalar fields
using the quadratic form 
\be
Q_A(\phi) = (\phi,A\phi)
\label{qaf}
\ee
associated to $A$, via \cite{ishihoso,stalker}
\be
||\phi||^2_{H^1} = (\phi,\phi) + Q_A(\phi)\;\;,
\label{sob}
\ee
thus enforcing the condition that the energy be finite.  For present
purposes we simply note that at least for the Friedrichs extension $A_F$
of $A$, based on the closure of the quadratic form $Q_A(\phi)$ with
respect to the $L^2$ norm, the resulting potential energy $Q_{A_F}(\phi)$
is finite (and positive) by definition without having to invoke Sobolev
spaces (see also the discussion in \cite{ishiwald,ishiwald2}).\footnote{Working 
with such a Sobolev space structure is certainly
possible but also complicates the determination of self-adjoint extensions
of $A$, since e.g.\ studying the closure of $A$ now involves studying the
sixth order operator $A^3$, arising from the term $||A\phi||_{H^1}^2 =
(A\phi,A\phi) + 
(A\phi, A^2\phi)$ in the operator norm.} We will use specifically this
extension in the discussion of section 3.5 below.

\subsection{Essential Self-Adjointness and the Horowitz-Marolf Criterion}

The spatial part $A$ of the wave operator is real and symmetric (with
respect to an appropriate scalar product on a $C_{0}^{\infty}$ domain
of $A$), and as such has self-adjoint extensions, each leading to a
well defined (and reasonable \cite{ishiwald}) time-evolution. If the
self-adjoint extension is not unique, however, i.e.\ if the operator
is not essentially self-adjoint, then also the corresponding
time-evolution is not uniquely determined. Thus the Horowitz-Marolf
criterion \cite{hm} (unique time-evolution without having to impose
boundary conditions at the singularity) amounts to the condition that
the operator $A$ be essentially self-adjoint.

To test for essential self-adjointness \cite{RS}, one can e.g.\ use
\cite{hm} the standard method of Neumann deficiency indices or the
Weyl limit point -- limit circle criterion (employed in this context
in \cite{khw}). Roughly speaking, in order for $A$ to be essentially
self-adjoint the (effective) potential $V_{\mathrm{eff},\ell}$ appearing
in the operator $\tilde{A}_{\ell}$ has to be sufficiently repulsive near
$x=0$ to prevent the waves $\tilde{\phi}$ from leaking into the singularity.

Concretely, in the present case, where we only have control over
the operator $A$ near the singularity at $x=0$, the criteria for
the operator $\tilde{A}_{\ell}$ to be essentially self adjoint on
$C_{0}^{\infty}(\RR_+)$ at $x=0$ boil down to the following elementary
conditions on the effective potential $W \equiv V_{\mathrm{eff},\ell}$
\cite{RS}:

\begin{itemize}
\item If 
\be
W(x) \geq \frac{3}{4} x^{-2}
\label{con1}
\ee
near zero, then $-\del_x^2 + W(x)$ is essentially self-adjoint at $x=0$. 
\item
If for some $\epsilon > 0$ 
\be
W(x) \leq \left(\frac{3}{4}-\epsilon\right) x^{-2}
\ee
(in particular also if $W(x)$ is decreasing) near $x=0$, then
$-\del_x^2 +W(x)$ is not essentially self-adjoint at $x=0$.
\end{itemize}

The significance of the factor $3/4$ can be appreciated by looking at the
critical (and relevant for us) case of an inverse square potential
\be
W(x) = s(s-1)x^{-2}\;\;.
\ee
In this case the leading behaviour of the two linearly independent solutions
of the equation 
\be
\left(-\del_x^2 + W(x)\right) 
\tilde{\phi}_{\lambda}(x) = \lambda \tilde{\phi}_{\lambda}(x)
\label{lambda}
\ee
near $x=0$ is given by the 
two linearly independent solutions of the equation
\be
\left(-\del_x^2 + W(x)\right) \tilde{\phi}_{0}(x) = 0\;\;,
\ee
namely 
\be
\tilde{\phi}_0 \sim x^s \;\;\;\;\;\;\mathrm{or}
\;\;\;\;\;\; \tilde{\phi}_0 \sim x^{1-s}
\ee
Thus both solutions are square integrable near $x=0$ when $2s > -1$ and 
$2(1-s) > -1$, or 
\be
-\frac{1}{2} < s < \frac{3}{2}\;\;\;\;\LRa \;\;\;\;s(s-1) < \frac{3}{4}\;\;,
\label{sineq}
\ee 
and in this range of $c=s(s-1)$ the potential is limit circle and the
self-adjoint extension is not unique. Conversely, it follows that for
$c\geq 3/4$ the solutions of equation (\ref{lambda}) for $\lambda =
\pm i$ (which are necessarily complex linear combinations of the two
linearly independent real solutions) are not square-integrable near $x=0$.
Thus the deficiency indices are zero and the operator is essentially
self-adjoint for $c \geq 3/4$.

Even when there are two normalisable solutions, all is not lost however,
as it may be indicative of the possibility (or even necessity)
to continue the fields and/or the metric through the singularity
\cite{gkos}. Evidently, such an analytic continuation requires some 
thought (to say the least) in the case of Szekeres-Iyer metrics with
generic (non-rational) values of $p$ and $q$.

\subsection{The Horowitz-Marolf Criterion for Power-Law Singularities}

In the case at hand, timelike singularities of power-law type, the
effective potential is given by (\ref{siveff}) with $\eta =-1$
and $s=qd/4$.  We had already seen in section
2.5 that the mass term is never dominant at $x=0$ and we can therefore also
set $m^2 =0$. Thus the operator of interest is
\bea
\tilde{A}_{\ell} &=& - \del_x^2 + V_{\mathrm{eff},\ell}(x)\non
V_{\mathrm{eff},\ell}(x)&=&s(s-1)x^{-2} + \ell_d^2 x^{p-q} \;\;,
\eea
It is now straightforward to determine for which values of $(p,q)$ the
classical singularities at $x=0$ become regular or remain singular when probed
by scalar waves. First of all, we will show that we can 
reduce the analysis to the case $\ell=0$:
\begin{itemize}
\item
For $q<p+2$, the first term in the potential is dominant and independent
of $\ell$. Thus $A$ is essentially self-adjoint iff $\tilde{A}_{\ell=0}$
is essentially self-adjoint. As we know from (\ref{con1}), this condition
is satisfied iff $s(s-1)\geq 3/4$.
\item For $q> p+2$, the operators $\tilde{A}_{\ell}$ for $\ell\neq 0$
are essentially self-adjoint by the criterion (\ref{con1}). Thus $A$ is
essentially self-adjoint iff $\tilde{A}_{\ell=0}$ is. 
\item In the borderline case $q=p+2$, for $\ell\neq 0$ we have
\be
\ell \neq 0 \;\;\;\;\Ra\;\;\;\;
s(s-1) + \ell_d^2 \geq 3/4 
\ee
(with equality only for $s=1/2$ and $\ell=d=1$). Even in this case, therefore,
all the $\tilde{A}_{\ell}$ with $\ell \neq 0$ are essentially self-adjoint
and only $\tilde{A}_{\ell=0}$ needs to be examined. 
\end{itemize}
We can thus conclude that the operator $A$ is essentially self-adjoint iff
$s(s-1) \geq 3/4$ and that, in view of (\ref{sineq}),
it fails to be essentially self-adjoint for
\be
A \;\;\mathrm{not\;\; e.s.a.}\;\;\;\;\LRa \;\;\;\;
-\frac{1}{2} < s < \frac{3}{2}\;\;\;\;\LRa\;\;\;\;
-\frac{2}{d} < q < \frac{6}{d}\;\;.
\label{esabound}
\ee

\subsection{The Significance of the (Strict) Dominant Energy Condition}

While this has been rather straightforward, one of the virtues of the
present approach, based on using a class of metrics appropriate for a
generic singularity of power-law type, is that it allows us to draw a
general conclusion regarding the relation between the Horowitz-Marolf
criterion and properties of the matter (stress-energy) content of the
space-time near the singularity. 

Indeed, as we will now show, whenever the matter content of the
near-singularity space-time is  sufficiently ``positive'' (in the sense
of the strict DEC, as it turns out), the space-time remains singular
according to the Horowitz-Marolf criterion, i.e.\ when probed with
scalar waves.

We can deduce from (\ref{gsi})
that metrics with timelike power-law singularities satisfying the strict
DEC lie in a bounded region inside the strip $0 < q < 2/d$ \cite{bbop2}. 
Indeed, for $q < p+2$ only the second terms in (\ref{gsi}) are relevant,
and one finds
\be
\rho -P_r = \trac{1}{4}dq(2-dq)\;x^{-(p+2)}\;\;.
\ee
Thus one has
\be
\rho -P_r > 0 \;\;\;\; \LRa \;\;\;\; 0 < q < \frac{2}{d}\;\;.
\label{tldec}
\ee
In particular, therefore, it follows from (\ref{esabound}) that for such
metrics the operator $A$ is not essentially self-adjoint and we can
draw the general conclusion that

\begin{quote}
metrics with timelike singularities of power-law type satisfying the strict
Dominant Energy Condition remain singular when probed with scalar waves.
\end{quote}

Even though metrics with $q=2/d$, say, like negative mass Schwarzschild,
still satisfy the bound (\ref{esabound}), thus remain singular while
obeying an extremal equation of state, we cannot strengthen the above
statement to include general metrics with extremal equations of state.
This can be seen e.g.\ from examples in \cite{hm} and is due to the fact
that extremal metrics can also be found elsewhere in the $(p,q)$-plane,
in particular in the region $q > p+2$, while violating the bound
(\ref{esabound}).

\subsection{The Friedrichs Extension and ``Hospitable'' Singularities}

In the previous section we have discussed self-adjoint extensions of
(the spatial part $A$ of) the Klein-Gordon operator. We have not discussed,
however, what these self-adjoint extensions imply about the behaviour
of the allowed scalar fields $\phi$ (those in the domain of the self-adjoint
extension of $A$) near the singularity at $x=0$. 

It is certainly possible that self-adjointness can be achieved by allowing
only scalar fields that vanish at the singularity. In some sense, then,
the singularity remains excluded from the space-time and is not probed
directly by the scalar field $\phi$. We will see that this is indeed what
happens in (in a precise sense) one half of the cases in which there is a
unique self-adjoint extension.

However, it is a priori also possible (and perhaps more interesting)
to have a well-defined time-evolution (which we take to mean ``defined
by some self-adjoint extension'' \cite{ishiwald}) with scalar fields
that are permitted to be non-zero at the singularity.  In that case,
the singularity would be probed more directly by the scalar field,
and one might then perhaps like to define a classical singularity to be
``hospitable'' (for a scalar field), if there is a self-adjoint extension
which allows the scalar fields to take non-zero values at the locus of
the singularity.  We will see that this possibility is indeed realised as
well, not only for the other half of the essentially self-adjoint cases,
but also for e.g.\ the Friedrichs extension $A_F$ of the operator $A$ in a
certain range of parameters for which $A$ is not essentially self-adjoint.

To address these issues, we need to determine the domain of definition
of the relevant self-adjoint extension of $\tilde{A}_{0} = -\del_x^2
+ c x^{-2}$ for $c=s(s-1)\in [-1/4,\infty)$. For $\tilde{A}_0$
essentially self-adjoint, i.e.\ $c \geq 3/4$, this can be done by
explicitly determining the domain of the closure $\bar{A}_0$ of the operator
$\tilde{A}_0$. While we have done this (see also \cite{horacio}), 
alternatively, for all $c \geq  -1/4$, one can determine the domain
of the Friedrichs extension $\tilde{A}_F$ 
of $\tilde{A}_{0}$, constructed from the
closure of the associated quadratic form. For $c \geq 3/4$, such 
that $\tilde{A}_0$ is essentially
self-adjoint, its unique self-adjoint extension of course agrees with
the Friedrichs extension. Precisely this problem has been addressed and
solved in \cite{fex}, and instead of reinventing the wheel here we can
draw on the results of that reference to analyse the issue at hand.

The main result of \cite{fex} of interest to us is their Theorem 6.4.
Applied to the operator $\tilde{A}_0$, this theorem\footnote{ Actually,
in \cite{fex} a more general operator, including in particular a
non-zero harmonic oscillator term $Bx^2$, was studied. However, this
term serves only to regularise the wave functions at infinity. Since we
are concerned with the behaviour at $x=0$, this term is of no consequence
for the present considerations.} states that the domain of the Friedrichs
extension $\tilde{A}_F$ of $\tilde{A}_0$ is
\bea
D(\tilde{A}_F) = \{f\in L^2(0,\infty):&& f(0)=0, f\in A(0,\infty), 
\del_x f \in L^2(0,\infty), \non
&&x^{-1}f \in L^2(0,\infty), (-\del_x^2 + c x^{-2})f\in
L^2(0,\infty)\}
\label{fdom}
\eea
where $A(0,\infty)$ denotes the space of absolutely continuous functions.
In \cite{fex}, this result was established for $c>0$. As far as we can see,
this result is correct, as it stands, also for $-1/4< c < 0$. 
We will comment on the special case $c=-1/4$ below.

We will now extract from this result some restrictions on the behaviour of 
$f$ near $x=0$ (assuming that we can model the leading behaviour of $f$ 
as $x\ra 0$ by some power of $x$):
\begin{enumerate}
\item From the condition 
$x^{-1}f\in L^2$ we learn that $f(x) \sim x^{\frac{1}{2} +
\epsilon}$ for some $\epsilon > 0$.
Then the conditions $f(0) = 0$ and $\del_x f \in L^2$ are also
satisfied.
\item The remaining condition $(-\del_x^2 + c x^{-2})f\in L^2$ can be
satisfied in one of two ways. Either both terms separately are in $L^2$
or $f$ lies in the kernel of the operator (as $x\ra 0$). In the former
case, we find the condition $f(x) \sim x^{\frac{3}{2}+\epsilon}$ with
$\epsilon > 0$. In the
latter case, since the two functions in the kernel are $x^s$ and $x^{1-s}$,
with (as usual) $c=s(s-1)$, we now need to distinguish several cases:
\begin{enumerate}
\item $c> 3/4$: this means that $s>3/2$ or $s < -1/2$. The solution $x^s$
with $s>3/2$, i.e.\ $f(x) \sim x^{\frac{3}{2}+ \epsilon}$,
yields nothing new. The solution $x^{1-s}$ with $s > 3/2$ 
(or, equivalently, the solution $x^s$ with $s< -1/2$) is ruled out by 
condition 1.
\item $c=3/4$: this means that $s=3/2$ or $s=-1/2$. In this case, we can 
allow $x^{3/2}$ and thus relax the domain to include functions $f(x) \sim 
x^{\frac{3}{2} + \epsilon}$, now with $\epsilon \geq 0$.
\item $-1/4 < c < 3/4$: thus $-1/2 < s < 3/2$ and $s\neq 1/2$. Thus
the solution $x^s$ is adjoined to the functions 
$\{x^{\frac{3}{2}+\epsilon}\}$ for $s> 1/2$, and the solution $x^{1-s}$
for $s< 1/2$. 
\end{enumerate}
\end{enumerate}

It remains to discuss the special value $c=-1/4$ or $s=1/2$ which is
not covered by the formulation of the domain in (\ref{fdom}).  This is
the minimal allowed value of interest to us ($c=s(s-1)$ with $s$ real),
and also the minimal value for which the operator remains positive (and
thus has a Friedrichs extension). In this case, the two solutions are
$x^s = x^{\frac{1}{2}}$ and $x^{\frac{1}{2}}\log x$, and we checked that,
as expected, the domain of the Friedrichs extension includes $x^{1/2}$.
This can also be deduced e.g.\ from \cite{kkpljp}, which moreover illustrates
nicely some of the weirdness of non-Friedrichs extensions. 

The above discussion shows that the two definitions (\ref{ef1}) and
(\ref{ef2}) of the energy, a priori differing by boundary terms
due to the integration by parts, agree for the Friedrichs extension
for $c> -1/4$ and differ only by a finite term for $c=-1/4$. The issue
of boundary terms for more general domains is discussed in 
\cite{gibharish}.

Returning to the original question of determining the behaviour of
the allowed scalar fields in the domain of the self-adjoint extension
of the spatial part $A$ of the Klein-Gordon operator, we need to now
undo the transformation $\phi \ra \tilde{\phi}$ from the initial scalar
fields $\phi$ to the half-densities $\tilde{\phi}$ that we performed in
section 2.1 to put $A$ into the form of a standard Schr\"odinger operator.

This transformation back from $\tilde{\phi}$ to $\phi$ is accomplished
by multiplication by $x^{-s}$. Now the upshot of the above discussion is
that the lowest power of $x$ appearing in the domain of $\tilde{A}_F$
is 
\be
\tilde{\phi}_{\mathrm{min}} \sim \left\{\begin{array}{cl}
x^{\frac{3}{2}+\epsilon}& \mathrm{for}\;\; s> 3/2\;\; \mathrm{or}
\;\;s < -1/2 \\
x^s& \mathrm{for}\;\; 1/2 \leq s \leq 3/2 \\
x^{1-s} & \mathrm{for}\;\; -1/2 \leq s \leq 1/2.  
\end{array}
\right.
\ee
Evidently
these functions are, in particular, positive powers of $x$. Thus they,
and therefore all functions in the domain, tend to zero for $x\ra 0$,
consistent with the condition $f(0)=0$ in (\ref{fdom}).  However this
is not necessarily true for the transformed functions, for which one
has ($\delta = \delta(s) > 0$ is a positive real number depending on $s$)
\be
\phi_{\mathrm{min}} = x^{-s}\tilde{\phi}_{\mathrm{min}} 
\sim \left\{\begin{array}{ll}
x^{\frac{3}{2}+\epsilon-s}= x^{-\delta} & \mathrm{for}\;\; s>
3/2\\
x^0=1& \mathrm{for}\;\; 1/2 \leq s \leq 3/2 \\
x^{1-2s} = x^{\delta}& \mathrm{for}\;\; -1/2 \leq s < 1/2\\
x^{\frac{3}{2}+\epsilon-s}=
x^{2+\delta} & \mathrm{for}\;\; s < -1/2
\end{array}
\right.
\ee
The final result is the simple statement that a $\phi$ in the domain of
the Friedrichs extension $A_F$ of $A$ necessarily goes to zero for $s <
1/2$, $\phi$ can be non-zero (but remains bounded) for $1/2 < s \leq 3/2$, 
and can become increasingly singular for large $s> 3/2$. 

Note that this statement is not invariant under $s\ra 1-s$. Indeed, while
the operator $-\del_x^2 + s(s-1)x^{-2}$ has this invariance, and therefore
also statements about its essential self-adjointness are symmetric under
$s\ra 1-s$ (as we have seen), the unitary transformation between $\phi$
and $\tilde{\phi}$ depends linearly on $s$ and thus leads to a behaviour of the
original scalar fields $\phi$ that does not have this symmetry.

Once again we find a pleasing relation with the DEC, since the watershed
happens exactly at $s=1/2 \LRa q=2/d$ which, as we have seen, corresponds
to $\rho = P_r$. Timelike singularities satisfying the strict DEC have
$0< q< 2/d$ (\ref{tldec}), thus $0 < s < 1/2$.  Moreover, metrics with
$s\leq -1/2$ have a unique self-adjoint extension $(c \geq 3/4$), thus
are regular in the Horowitz-Marolf sense, but are not ``hospitable''
in the sense described above, while those with $s\geq 3/2$ are.

\subsection*{Acknowledgements}

MB is grateful to Martin O'Loughlin for collaboration on these issues in
the early stages of this work \cite{mbmo}
and for numerous discussions. This work
has been supported by the Swiss National Science Foundation and by the
EU under contract MRTN-CT-2004-005104.

\rnc{\Large}{\normalsize}

\end{document}